\documentclass[conference]{IEEEtran}
\IEEEoverridecommandlockouts
\usepackage{fancyhdr}
\fancypagestyle{mahmood}{%
  \fancyhf{} % clear all fields
  
  \fancyhead[C]{\footnotesize \textcopyright 2022 IEEE. Personal use of this material is permitted.  Permission from IEEE must be obtained for all other uses, in any current or future media, including reprinting/republishing this material for advertising or promotional purposes, creating new collective works, for resale or redistribution to servers or lists, or reuse of any copyrighted component of this work in other works.}
}%
\makeatletter
\let\ps@IEEEtitlepagestyle\ps@mahmood
\makeatother
\usepackage{cite}
\usepackage{amsmath,amssymb,amsfonts}
\usepackage{algorithmic}
\usepackage{graphicx}
\usepackage{textcomp}
\usepackage{xcolor}
\usepackage{pgfplots} 
\usepgfplotslibrary{fillbetween}
\usepackage{subcaption}
\usepackage{float}

\usepackage{tikz}
\usepackage{tkz-euclide}
\usetikzlibrary{calc,intersections,through,backgrounds,snakes}
\usepackage{pgfplots}
\pgfplotsset{compat=1.8}
\usepgfplotslibrary{statistics}
\usepackage{bchart}
\usetikzlibrary{positioning}

\usepackage{booktabs}
\usepackage{threeparttable}
\usepackage{multirow}

\usepackage{siunitx}
\def\BibTeX{{\rm B\kern-.05em{\sc i\kern-.025em b}\kern-.08em
    T\kern-.1667em\lower.7ex\hbox{E}\kern-.125emX}}

\begin{document}
\bstctlcite{IEEEexample:BSTcontrol}
% \title{Spiking Neural Network Decoder for Implantable Brain-Machine Interface\\
% \title{Efficient Spiking Neural Network for Velocity Regression in Implantable Brain--Machine Interface
\title{An Energy-Efficient Spiking Neural Network for Finger Velocity Decoding for Implantable Brain-Machine Interface
%Finger Velocity Regression using Efficient Spiking Neural Network for Implantable Brain--Machine Interfaces
}

% \and
% \and
% \and
% \and
% \and
% \author{\IEEEauthorblockN{authors removed}}
\author{\IEEEauthorblockN{Jiawei Liao$^{*\dag}$, Lars Widmer$^\dag$, Xiaying Wang$^\dag$, Alfio Di Mauro$^\dag$, Samuel R. Nason-Tomaszewski$^\ddag$, \\ Cynthia A. Chestek$^\ddag$, Luca Benini$^\dag$, Taekwang Jang$^\dag$}
\IEEEauthorblockA{ETH Zurich, Switzerland$^\dag$  \  University of Michigan, USA$^\ddag$  \ *Email: liaoj@iis.ee.ethz.ch}}
% \thanks{* Corresponding author: liaoj@iis.ee.ethz.ch}

\maketitle

\begin{abstract}
% This document is a model and instructions for \LaTeX.
% This and the IEEEtran.cls file define the components of your paper [title, text, heads, etc.]. *CRITICAL: Do Not Use Symbols, Special Characters, Footnotes, 
% or Math in Paper Title or Abstract.
Brain-machine interfaces (BMIs) are promising for motor rehabilitation and mobility augmentation. High-accuracy and low-power algorithms are required to achieve implantable BMI systems. In this paper, we propose a novel spiking neural network (SNN) decoder for implantable BMI regression tasks. The SNN is trained with enhanced spatio-temporal backpropagation to fully leverage its ability in handling temporal problems. The proposed SNN decoder achieves the same level of correlation coefficient as the state-of-the-art ANN decoder in offline finger velocity decoding tasks, while it requires only 6.8\% of the computation operations and 9.4\% of the memory access. 
% Also, the computation complexity analysis has been performed to verify the efficiency of the proposed decoder.
% The proposed SNN shows its feasibility in achieving power-efficient hardware implementation for close-loop control tasks in the future.
\end{abstract}

% This document is a model and instructions for \LaTeX.
% Please observe the conference page limits. 
\section{Introduction}
Brain-machine interfaces (BMIs) acquire signals generated by neurons, use them to decode users' intentions, and convert them into commands that can be used to control actuators like robotic arms and prostheses. They are essential tools in motor rehabilitation by restoring the movement ability of amputees and tetraplegic patients \cite{ajiboye_restoration_2017,collinger_7_2013,hochberg_reach_2012,hotson_individual_2016}.
% in augmenting users' mobility and especially in motor rehabilitation by restoring the movement ability of amputees and patients suffering from High cervical spinal cord injury. % Such brain interface systems also help patients suffering from neurological disorders. 

To achieve the objective, it is essential to develop accurate algorithms to decode motor function from neural signals. Linear decoders, including linear regression \cite{collinger_7_2013}, linear discriminant analysis \cite{hotson_individual_2016}, and variants of Kalman filters \cite{ajiboye_restoration_2017,nason_real-time_2020,hochberg_reach_2012} have been developed to perform arm and hand control. To further improve the accuracy, nonlinear decoders such as recurrent neural network \cite{sussillo_recurrent_2012,hosman_bci_2019} and feed forward neural network \cite{glaser_machine_2020,willsey_real-time_2021} are actively investigated. %by exploiting the potential nonlinear relationship between neural activities and movements.
While neural networks are powerful, they come at the cost of high computational complexity, implying high energy consumption in hardware implementation \cite{sze_efficient_2017}.
% Nonlinear: RNN\cite{sussillo_recurrent_2012,hosman_bci_2019}.
% A neural decoder based on Kalman filter (KF) In~\cite{nason_real-time_2020} has demonstrated its good performance in decoding movements of two finger groups during an online control experiment. 
% Although KF is highly power efficient, it is a linear predictor. To capture the non-linear features, researchers have developed a neural network decoder and shown its superior performance than KF~\cite{willsey_real-time_2021}.
%Unlike the common wearable surface EEG systems, implantable BMI systems can acquire signals with higher signal-to-noise ratio for more accurate prediction. 
For BMIs, that perform their inference tasks in the implanted devices to reduce the wireless communication overhead for the large volumes of raw data, their power efficiency is a critical concern to limit the heat generation and to protect the surrounding tissues. As a result, highly energy efficient approaches, that match the accuracy achieved by artificial neural network (ANN) should be designed.
%However, it also imposes more stringent power consumption requirement. While neural networks can be a powerful decoder, it comes with large amount of computation, indicating high power consumption in hardware implementation. 

Spiking neural networks (SNNs), a computing paradigm inspired by biological neural networks, have potential for achieving energy-efficient computation by leveraging sparsity introduced by the asynchronous feature of the neurons \cite{10.3389/fnins.2018.00774}. 
While SNNs have been heavily studied to solve classification tasks~\cite{wu_spatio-temporal_2018,zheng_going_2020}, they are perfect candidates for solving regression tasks with spatio-temporal inputs and outputs, such as motor decoding, thanks to their nature of taking time series as input and generating time series output.
% Most of the existing studies using SNNs focus on solving classification tasks, especially image classification~\cite{wu_spatio-temporal_2018,zheng_going_2020}. 
% SNNs naturally take time series as input and they generate time series outputs, making them a perfect candidate for solving regression tasks with spatio-temporal inputs and outputs, such as motor decoding.

There are three main ways to construct SNNs: 1) 
% ANN-to-SNN conversion is a common method, which requires training of ANN before converting it to the SNN.
The first is ANN-to-SNN conversion. This requires an ANN trained before being converted to the SNN. 
Typically, it relies on rate coding and tries to mimic the computation of the ANN by setting the parameters of the SNN to correlate the spiking neurons' firing rates with the neurons' activation of the original ANN~\cite{rueckauer_conversion_2017}. This conversion approach has shown similar performance as its ANN counterparts.
%, but it typically needs a high number of time steps and spike counts for rate approximation. 
% In the same vein, \cite{dethier_spiking_2011} introduced an SNN decoder converted from a Kalman filter decoder. The converted SNN achieves similar performance as its Kalman filter counterpart, but high spike counts are required due to the conversion process and rate approximation.
% , which leads to higher spike count, limiting the energy efficiency, and lower potentials of achieving better performance. 
2) The second is unsupervised learning such as spike-timing-dependent-plasticity (STDP). This is a biologically plausible approach relying on local learning rules~\cite{masquelier_unsupervised_2007}. However, without global supervision, so far, it has not achieved state-of-the-art performance in terms of accuracy and energy efficiency. 3) The third is SNN backpropagation such as spatio-temporal-back-propogation (STBP)~\cite{wu_spatio-temporal_2018,zheng_going_2020}. By applying a surrogate function in the backward flow to approximate the derivative of the spike activity, the error backpropagation path can be established for gradient descent training. This approach utilizes the temporal feature of the input and has demonstrated good performance in classification tasks. In this work, we construct the SNN using the SNN backpropagation method.

% \cite{stdp bmi} presents an SNN for muscle activity and kinematics decoding with EEG signals using 'NeuCube' brain-inspired SNN architecture trained with both STDP-based unsupervised training and supervised training. It provides better interpretability than other neural network models. However, it uses highly complex network model with redundancies mimicking brain, which is not suitable for low-power hardware implementation. 
% In~\cite{gehrig_event-based_2020}, authors train their SNN using SNN backprogation training with Spike LAYer Error Reassignment (SLAYER) framework~\cite{shrestha_slayer_2018} and successfully apply the SNN to predict camera's angular velocity~\cite{gehrig_event-based_2020}. However, it employs a complex neuron model Spike Response Model, which is not hardware-friendly. Additionally, the paper shows the successful prediction for 100ms chunks but it has not demonstrated its performance in long-term continuous dataset, which is essential for real-time prediction in real world. \todo{rephrase this paragraph and include STDP BMI}

% In~\cite{gehrig_event-based_2020}, authors present a SNN trained with Spike Layer Error Reassignment (SLAYER) framework for camera's angular velocity prediction. However, it employs a less hardware-friendly Spike Response neuron model. The results are demonstrated in
In this paper, we propose a novel low-complexity SNN trained with improved STBP backpropagation method to predict the open-loop two-finger velocities for implantable BMIs. The main contributions are:
\begin{itemize}
    \item We propose and open source\footnote{https://github.com/liaoRichard/SNN-for-Finger-Velocity-iBMI} a novel low-complexity SNN based on leaky-integrated-and-fire (LIF) neurons for regressing continuous-time finger velocity decoding for low-power implantable BMIs. 
    \item We demonstrate that the STBP SNN training strategy can be enhanced with techniques such as the neuron reset-by-subtraction scheme and trainable decay factors to achieve the same level of performance as a state-of-the-art ANN decoder.
    \item We validate the accuracy of the proposed SNN on two datasets recorded from non-human primates' hand area of primary motor cortex for open-loop decoding tasks with two individual finger groups, in a streaming fashion, to mimic a real-time decoding scenario. 
    \item We perform an SNN computation complexity analysis and compare it with both the state-of-the-art ANN decoder and the ANN-converted SNN. The proposed SNN requires only $6.8\%$ computation operations and $9.4\%$ memory access compared to the ANN, indicating its potential for more energy-efficient hardware implementation.

\end{itemize}
% We also provide our source code. 

\vspace{-5pt}
\section{Methods}
\vspace{-5pt}
% In this section, we present the proposed SNN implementation and its training method.

\subsection{Neuron model} \label{sec:neuron_model}
% \vspace{-3pt}
We use the adjusted LIF neuron model for its simplicity in hardware implementation~\cite{izhikevich_which_2004}. As demonstrated in \cite{izhikevich_which_2004}, the LIF neuron model shows a good trade-off between cognitive capabilities and computational complexity, making it a suitable candidate for the implementation on embedded platforms with constrained-resources. The implemented neuron model is described in  \eqref{eq:mem_upd}, \eqref{eq:inp_current}, and \eqref{eq:spike_fun}. 
\vspace{-2pt}
\begin{equation}
\label{eq:mem_upd}
u^{l}(t) = \tau (u^{l}(t-1)-s^{l}(t-1)V_{th}) + I^{l}(t) 
\vspace{-4pt}
\end{equation}
\begin{equation}
\label{eq:inp_current}
I(t) = \sum_{i=0}^{M-1} w^{l} s^{l-1}(t)
\vspace{-4pt}
\end{equation}
\begin{equation}
\label{eq:spike_fun}
s^{l}(t) = \begin{cases}
1   & u^{l}(t) \ge V_{th} \\
0   & u^{l}(t) < V_{th}
\end{cases}
\vspace{-4pt}
\end{equation}
In \eqref{eq:mem_upd}, $u^{l}$ represents the membrane potential in $l$th layer and $I$ represents the input current expressed by \eqref{eq:inp_current}. $V_{th}$ denotes the threshold voltage of the neuron. M is the number of input connections of the neuron.  $s$ is the status of the spike defined by \eqref{eq:spike_fun} and $s$ equals one when the neuron fires. $\tau$ is the decay factor of the leaky neuron. The LIF integrates the input current and leaks at a rate of $\tau$. When the membrane potential exceeds the threshold voltage, the neuron fires and the membrane potential decreases by $V_{th}$. Inspired by \cite{tan_improved_2021}, we apply reset-by-subtract scheme instead of reset-by-zero scheme to avoid loss due to the over adjustment after spiking. 
% In our experiment, we have found that with reset-by-subtract scheme, the accuracy increases compare to reset-by-zero scheme.

% \begin{figure*}[htb]
%  \centering\includegraphics[width=0.8\linewidth]{figures/network.pdf}
%  \caption{SNN in inference and unfolded for training}
%  \label{fig:network_arch}
%  \vspace{-10pt}
% \end{figure*}

\begin{figure}[t]
 \centering\includegraphics[width=\linewidth]{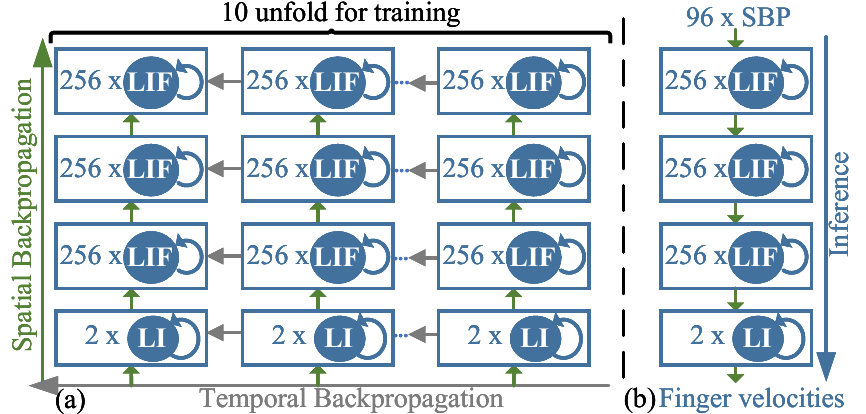}
 \caption{Proposed SNN unfolded for training (a) and in inference (b)}
 \label{fig:network_arch}
 \vspace{-15pt}
\end{figure}

% \vspace{-5pt}
\subsection{Network architecture}
% \vspace{-3pt}
We use spiking band power (SBP) proposed in \cite{nason_low-power_2020} as the input to the network. SBP is a neural feature for motor prediction defined as an average of absolute $\SI{300}-\SI{1000}{\hertz}$ band-pass-filtered signal. The proposed network handles the SBP features averaged in time frames from 96 channels. 

Our proposed SNN architecture is depicted in Fig.~\ref{fig:network_arch}. It consists of four fully connected layers inspired by \cite{willsey_real-time_2021}. Unlike \cite{willsey_real-time_2021}, where the authors use a convolutional layer as the input layer to capture the temporal information from multiple time frames, our network directly receives the features from a single time frame as input.
This choice has been made to exploit an intrinsic property of SNNs, i.e., the membrane potential of each neuron acts as a memory element, and stores temporal information from previous inputs. 
% This allow us to better exploit an intrinsic property of SNNs, i.e., the membrane potential of each neuron acts as a memory element, and stores information from previous inputs. 
% The SNN trained with STBP can leverage the spatio-temporal feature \cite{wu_spatio-temporal_2018}.
% In other words, an SNN trained with spatio-temporal-backpropagation can extract temporal features by itself.
The output layer uses two non-spiking neurons that follow the membrane equation $u(t) = \tau u(t-1) + I(t) $. The membrane voltages $u(t)$ are then taken directly as the continuous-valued outputs for the prediction of two finger velocities for every input. 

%The proposed SNN consists of four fully connected layer. The input layer takes input features from 96 channels and it handles data from one time frame at a time. The output layer uses two non-spiking neuron described in \eqref{eq:non_spike_neuron} to produce continuous value output for the prediction of two finger velocities for every input. 
Except for the first layer of the network whose input current is the weighted sum of the SBP features, in all hidden layers, the total input current $I$ of each LIF neuron is obtained as the weighted sum of the input spikes whose values are either 1 or 0, as described in \eqref{eq:inp_current} and \eqref{eq:spike_fun}. Therefore, one of the advantages of using an SNN instead of a conventional ANN is that the input current $I$ can be calculated as additions of weights instead of Multiply-Accumulate (MAC) operations thanks to the 1-bit binary input spikes. Except for the last output layer, all hidden neurons fire only when the membrane potentials exceed the threshold, introducing a high degree of sparsity in the intermediate features. The reduced computational complexity as well as the high degree of sparsity are the two main sources of efficiency of the proposed SNN compared to a conventional ANN.

Between successive layers, the batch normalization is implemented to improve convergence. To handle the extra temporal dimension in SNN and to avoid both gradient vanishing and gradient explosion problems by balancing the firing rate, we use the special threshold-based batch normalization introduced in \cite{zheng_going_2020}. 
% Taking the value of the threshold into account helps to balance the firing rate and avoid both gradient vanishing and gradient explosion problems for deep networks.
Dropout is implemented for regularization during training. The dropout is performed for only spatial dimensions. At each time step, a new dropout mask is generated randomly. 
\vspace{-10pt}
\subsection{Training methods} \label{sec:train_method}
% \vspace{-3pt}
Spiking functions for neurons in SNNs are not directly differentiable, so this has been an obstacle for direct backpropagation for SNNs for a long time. As suggested by \cite{shrestha_slayer_2018,wu_spatio-temporal_2018}, using a surrogate function to approximate the derivative of spike activity allows the gradient to propagate back through the neurons. In this work, we use a square surrogate function defined by \eqref{eq:surrogate_fun}.
\vspace{-4pt}
\begin{equation}
\label{eq:surrogate_fun}
\frac{\partial a}{\partial u} = \begin{cases}
1  &  \text{if} |u(t) - V_{th}| < 0.5 \\
0  &  \text{else}
\end{cases}
\vspace{-4pt}
\end{equation}

The training process is based on the publicly available PyTorch implementation of the STBP training method introduced in \cite{wu_spatio-temporal_2018}. The support for reset-by-subtract scheme, trainable decay factor, as well as non-spiking output layers were implemented additionally to the training framework.
% \footnote{https://github.com/ZLkanyo009/STBP-train-and-compression} 
Instead of defining a decaying factor $\tau$ as a hyperparameter shared by all the neurons, we make it a trainable parameter for each neuron, and optimize them at training time; the value for $\tau$ is clamped between 0 and 1 during the forward pass. With different decaying factors, each neuron has an independent behavior when receiving the input spikes, thereby showing a different sensitivity to the events received in the past; this feature increases the expressiveness of the SNN ~\cite{fang_incorporating_nodate,yin_accurate_2021}.

% In the results section we reported the performance of our SNN with and without the trainable tau. 

Fig.~\ref{fig:network_arch}a shows the unfolded network during the training process. The STBP allows backpropagation through both temporal and spatial dimensions. The network is unfolded for ten time frames during the training process. We experimentally observed that using more than ten frames does not improve the accuracy of the network, therefore, we used ten time-frames as the upper bound of our exploration. Additionally, we discarded the first two frames in the loss calculation,  because the network has not yet converged to a stable prediction. The loss is then backpropagated through spatial and temporal dimension in this unfolded network. By using a sliding window whose length is 10 time frames and overlap is 9 time frames, training samples are generated. These samples containing 10 time frames are then shuffled for training.

During the inference process, the network is not unfolded, and all neurons maintain and update their internal state through the whole process. One time frame is used once per inference. This operating scenario mimics the situation where a real-time prediction task runs on data that are streamed continuously to the network.
We applied AdamW optimizer \cite{loshchilov_decoupled_2018} with learning rate and weight decay of $2\times10^{-3}$ and $1\times10^{-2}$, respectively. The batch size used during training is 128. The membrane threshold $V_{th}$ is set to $0.4$. These hyperparameters are determined by grid search. 
%
% \subsection{Loss function and evaluation metrics}
Similarly to \cite{willsey_real-time_2021}, we use the time-integrated mean square error as the loss function during training while Pearson correlation coefficient is used as the metric to evaluate the performance of SNN as it is commonly used for neural decoding algorithm comparison \cite{nason_low-power_2020,willsey_real-time_2021}.

% \begin{equation}
% \label{eq:corrcoef}
% r = \frac{\sum (vel_{pred} - \overline{vel_{pred}})(vel_{true} - \overline{vel_{true}})}{\sqrt{\sum (vel_{pred} - \overline{vel_{pred}})^2 \sum(vel_{true} - \overline{vel_{true}})^2}}
% \end{equation}
\vspace{-2pt}
\section{Results and discussion}
\vspace{-5pt}
\subsection{Dataset}
\vspace{-3pt}
We evaluated the proposed SNN on two datasets recorded from non-human primates while they were performing two-degree-of-freedom finger tasks. The datasets contain positions, velocities of two fingers, and the SBP features from 96 channels.
% with a data rate of $\SI{1}{kHz}$. 
% The data are sampled every $\SI{1}{ms}$. 
Dataset A, also used in~\cite{willsey_real-time_2021}, contains $\SI{817}{s}$ data. Dataset B is the open-source dataset also used for~\cite{nason_real-time_2020}, containing $\SI{610}{s}$ data.
% Dataset A, also used in~\cite{willsey_real-time_2021}, contains data generated from $\SI{817}{s}$ recording. Dataset B is the open-source dataset, also used for~\cite{nason_real-time_2020}, containing data generated from $\SI{610}{s}$ recording.
The first $80\%$ of the data are used for training, the remaining $20\%$ are for validation. The SNN is trained for 24 epochs and 23 epochs for the evaluation on Dataset A and B respectively. The velocities and SBP features are averaged within time frames to be processed. Time frames' sizes are chosen to be $\SI{50}{ms}$ and $\SI{32}{ms}$ as introduced in \cite{willsey_real-time_2021,nason_real-time_2020}. Averaged SBP features are standardized by removing means and scaling to standard deviation of 1 before being fed into the SNN. Predicted velocity is also standardized with statistics from training set. Standardized data are recovered to original scale for inference. An inference is performed for every time frame to generate two finger velocities in a streaming fashion.

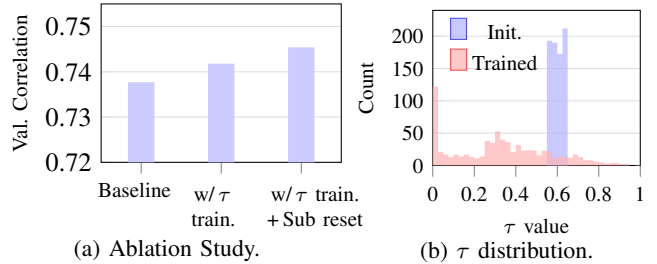
\begin{figure}[t]
        \centering
        \begin{subfigure}[t]{.49\linewidth}
            \centering
        \begin{tikzpicture}
            \begin{axis} 
            [ybar,
            width=1.1\linewidth,
            height=0.85\linewidth,
       ylabel = {\footnotesize Val. Correlation},
      ylabel near ticks,
      xlabel near ticks,
      tick pos=left,
      ymajorgrids,
            grid style={gray!25},
      ytick style={gray!25},
    xtick = {0,1,2},
    xticklabels = {\hspace{-5pt}{\footnotesize Baseline}, {\renewcommand{\arraystretch}{0.8}\begin{tabular}{c} {\footnotesize w/\,$\tau$\hspace{0.6em}{\color{white}.}} \\ \hspace{1pt}\footnotesize train.\hspace{0.6em}{\color{white}.} \end{tabular}},
    {\renewcommand{\arraystretch}{0.8}\begin{tabular}{c}\hspace{7pt} \footnotesize w/\,$\tau$ train.\\  \hspace{7pt} \footnotesize +\,Sub reset \end{tabular}}},
      ytick = {0.70, 0.71, 0.72, 0.73, 0.74, 0.75},
      yticklabels = {0.70, 0.71, 0.72, 0.73, 0.74, 0.75},
    %   label style={font=\small},
    %   tick label style={font=\small},
      xmin = -0.5, xmax = 2.5,
      ymin = 0.72, ymax = 0.755,
      %axis y line* = right
      legend style={
        at={(0.99, 0.99)},
        anchor=north east,
        draw=none,
        %/tikz/every even column/.append style={column sep=6pt},
        % /tikz/column 2/.style={column sep=10pt},
        % /tikz/column 4/.style={column sep=10pt},
        font={\footnotesize}
      },
            ]
\addplot [fill, color=blue!20,draw=none] coordinates { % fill=blue to set color
    (2,0.7454) 
    (1,0.7418) 
    %(2,0.678) 
    (0,0.7377)
};
% \node[anchor=south] at (axis cs: 2,0.7454){{\renewcommand{\arraystretch}{0.6}\begin{tabular}{c}\scriptsize +Sub\\ \scriptsize reset \end{tabular}}};
% \node[anchor=south] at (axis cs: 1,0.7418){{\renewcommand{\arraystretch}{0.6}\begin{tabular}{c}\scriptsize + Tau \\ \scriptsize trainable \end{tabular}}};
% \node[anchor=south] at (axis cs: 0,0.7189){{\renewcommand{\arraystretch}{0.6}\begin{tabular}{c}\scriptsize Baseline \end{tabular}}};
\end{axis}
        \end{tikzpicture}
        \vspace{-20pt}
        \caption{Ablation Study.}
        % \vspace{-5pt}
        \label{fig:abl_study}
    \end{subfigure}
    \hfill%        
    % here was an empty line which caused that the plots where not next
        % to each other but on top of each other
    \begin{subfigure}[t]{.49\linewidth}
            \centering
        \begin{tikzpicture}
            \begin{axis}
            [
            ybar,
            width= \linewidth,
            height=0.84\linewidth,
       xlabel = {$\tau$ value},
      ylabel = {Count},
      ylabel near ticks,
      xlabel near ticks,
      xtick = {0,0.2,0.4,0.6,0.8,1},
      xticklabels = {0,0.2,0.4,0.6,0.8,1},
      ytick = {0,50,...,240},
      yticklabels = {0,50,...,240},
      label style={font=\footnotesize},
    tick label style={font=\footnotesize},
      xmin = 0, xmax = 1,
      ymin = 0, ymax = 240,
      tick pos=left,
      %axis y line* = right
      ymajorgrids,
            grid style={gray!25},
      ytick style={gray!25},
      legend style={
        at={(0.57, 0.99)},
        anchor=north east,
        draw=none,
        fill=none,
        %/tikz/every even column/.append style={column sep=6pt},
        % /tikz/column 2/.style={column sep=10pt},
        % /tikz/column 4/.style={column sep=10pt},
        font={\footnotesize}
      },
            ]
\addplot+ [
        fill,
        draw=none,
        opacity=0.7,
        bar width=2pt,
        hist={bins=40}] %, data max=1,data min=0
        table [x = trained, y = init, col sep=comma] {figures/plot1_data_.csv};
\addlegendentry{Init.}
\addplot+ [
        fill,
        draw=none,
        opacity=0.7,
        bar width=2pt,
        hist={bins=40}] %, data max=1,data min=0
        table [x = init, y = trained, col sep=comma] {figures/plot1_data_.csv};
\addlegendentry{Trained}
\end{axis}
        \end{tikzpicture}
        \vspace{-7pt}
        \caption{$\tau$ distribution.}
        % \vspace{-5pt}
        \label{fig:tau_distr}
    \end{subfigure}%
\vspace{-3pt}
\caption{Ablation study and $\tau$ distribution}
\vspace{-3pt}
\label{fig:tau_distr_abl_study}
\vspace{-8pt}
\end{figure}

% \vspace{-5pt}
\subsection{Performance comparison}
% \vspace{-3pt}
The improvements that are achieved by using the proposed trainable decay factor and reset-by-subtract are quantified by the ablation study performed with dataset A and the results are depicted in Fig.~\ref{fig:abl_study}. %Degraded performance is shown with neither trainable decay factor nor reset-by-subtract scheme. 
Trainable decay factor and reset-by-subtract scheme jointly improve the correlation coefficient to over 0.74. Fig.~\ref{fig:tau_distr} shows the decay factors' distribution before and after training for each layer. %The decay factors show a wide spread after training which gives the neurons different dynamics.
%The wide spread of the decay factor suggest that various neuron dynamics may improve the expressiveness of the network.
The trained decay factors cover a wide range of values, allowing different neuron dynamics, which may improve the expressiveness of the network, as suggested by ~\cite{fang_incorporating_nodate}.

The results from our SNN experiments are summarized in Table \ref{tab:corr}. For comparison, we reproduced the KF predictor\cite{nason_real-time_2020} and the NN predictor\cite{willsey_real-time_2021} using the same parameters reported in the original papers. We also compare with the SNN converted from the ANN using SNN toolbox~\cite{rueckauer_conversion_2017}. The proposed network reaches better correlation coefficients than linear KF and achieves the same level of performance as the state-of-the-art ANN decoder.
The velocity predicted by the proposed SNN, ANN, and the true velocity of one of the fingers are plotted in Fig.~\ref{fig:vel_curve} and it shows a good match.
% \begin{figure}[htbp]
% \centerline{\includegraphics[width=0.98\linewidth]{figures/tau.png}}
% \caption{Tau distribution}
% \label{fig:tau}
% \end{figure}
% \begin{figure}[htbp]
% \centerline{\includegraphics[width=0.4\linewidth]{figures/ablation.png}}
% \caption{Ablation study}
% \label{fig:ablation}
% \end{figure}

% \begin{table}[b]
% \caption{Performance Comparison}
% \begin{center}
% \begin{tabular}{|c|c|c|c|c|}
% \hline
% \textbf{Data}&\multicolumn{4}{|c|}{\textbf{Decoder}} \\
% \cline{2-5} 
% \textbf{set} & \textbf{\textit{KF} \cite{nason_real-time_2020}}& \textbf{\textit{ANN} \cite{willsey_real-time_2021}}& \textbf{\textit{converted SNN}}& \textbf{\textit{Proposed SNN}} \\
% \hline
% A & 0.601 & 0.724 $^{\mathrm{a}}$ & 0.740 & 0.741 \\
% \hline
% B & 0.459 & 0.582 & 0.590 & 0.570 \\
% \hline
% \multicolumn{5}{l}{$^{\mathrm{a}}$ Reproduced according to papers} \\
% % \multicolumn{5}{l}\todo{direct conversion accuracy is correct, but is only a single conversion (converted an ANN that is close to mean accuracy, but still...} \\
% \end{tabular}
% \label{tab:corr}
% \end{center}
% \end{table}
%\vspace{-12pt}
%\vspace{-\abovedisplayskip}
  \noindent\begin{table}[b]
    \begin{minipage}[t]{.43\columnwidth}
\caption{Performance \\ comparison.}
% \centering
\label{tab:corr}
\vspace{-6pt}
\setlength{\tabcolsep}{2.6pt}
\renewcommand{\arraystretch}{1}
\small
% \centering
\begin{threeparttable}
%\begin{tabular}{@{}l @{\hspace{0.7\tabcolsep}} r @{\hspace{0.7\tabcolsep}} r@{}}
\begin{tabular}{@{}llr@{}}
\toprule
Dataset       & A     & B     \\ \midrule
KF \cite{nason_real-time_2020} & 0.601 & 0.459 \\
ANN \cite{willsey_real-time_2021} & 0.724$^\dagger$ & 0.582 \\
Converted & \multirow{2}{*}{0.740} & \multirow{2}{*}{0.590} \\
SNN & & \\
\textbf{Proposed} & \multirow{2}{*}{\textbf{0.745}} & \multirow{2}{*}{\textbf{0.582}} \\
\textbf{SNN} & & \\
%Converted SNN & 0.740 & 0.590 \\
%\textbf{Proposed SNN}  & \textbf{0.745} & \textbf{0.582} \\ 
\bottomrule
\end{tabular}
  \begin{tablenotes}
    \footnotesize
    \item $^\dagger$ Reproduced.
  \end{tablenotes}
    \end{threeparttable}
    \end{minipage}
    \hfill
\begin{minipage}[t]{.56\columnwidth}
      
\caption{Operation \\ comparison.}
% \centering
\label{tab:comp}
\vspace{-6pt}
\setlength{\tabcolsep}{2.8pt}
\renewcommand{\arraystretch}{1.2}
\small
% \centering
%\begin{tabular}{@{}l @{\hspace{0.3\tabcolsep}} r @{\hspace{0.6\tabcolsep}} r @{\hspace{0.6\tabcolsep}} r@{}}
\begin{threeparttable}
\begin{tabular}{@{}lrrr@{}}
\toprule
        %   & \multicolumn{2}{r}{Decoder} \\ \cmidrule(l){2-3} 
\multirow{2}{*}{Operation} & ANN    &Convert.       &  \textbf{Prop.}     \\
& \cite{willsey_real-time_2021} & SNN & \textbf{SNN}\\
\midrule
MAC       & 529K  & 5K        & \textbf{25K}         \\
ADD       & -     & 865K        & \textbf{33K}         \\
\hline
Total ops $^\dagger$    & 529K      & 293K    & \textbf{36K}        \\
\hline
Mem access      & 2116K    & 2615K    & \textbf{199K}        \\ \bottomrule
\end{tabular}
      \begin{tablenotes}
    \footnotesize
    \item $^\dagger$ Total MAC operations. 3 ADDs count as 1 MAC~\cite{horowitz_11_2014}.
  \end{tablenotes}
\end{threeparttable}
    \end{minipage}
  \end{table}
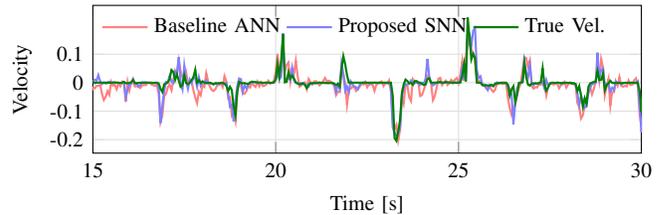
\begin{figure}[t]
  \centering
  \resizebox{\columnwidth}{!}{%
  \begin{tikzpicture}
    \begin{axis}[
      no markers,
      grid,
      grid style={gray!25},
      ytick style={gray!25},
      % ymode=log,
      width=\columnwidth,
      height=0.4\columnwidth,
      legend pos=south west,
      xlabel={ Time [s]},
      ylabel={ Velocity},
      xmin = 300, xmax = 600,
      xtick = {300,400,500,600},
      xticklabels = {15,20,25,30},
      ytick = {-0.2,-0.1,0,0.1},
      yticklabels = {-0.2,-0.1,0,0.1},
      label style={font=\footnotesize},
    tick label style={font=\footnotesize},
      legend style={
      at={(0,0.99)},
      anchor=north west,
        draw=none,
        fill=none,
        legend columns=-1,
        font={\footnotesize}
      },
      ]
     \addplot+ [thick,red!50!white] table [x=t, y=e, col sep=comma] {figures/plot4_data_.csv};
      \addplot+ [thick,blue!50!white] table [x=t, y=c, col sep=comma] {figures/plot4_data_.csv}; 
      \addplot+ [thick,green!50!black] table [x=t, y=a, col sep=comma] {figures/plot4_data_.csv};  
      
      \legend{Baseline ANN, Proposed SNN, True Vel.};
      
    \end{axis}
  \end{tikzpicture}
  }
  \vspace{-0.7cm}
  \caption{Example of predicted and true velocities.}
  \label{fig:vel_curve}
  \vspace{-17pt}
\end{figure}

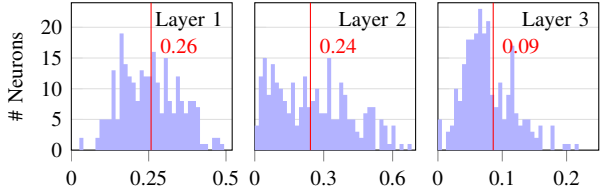
\begin{figure}[t]
\centering

% \begin{subfigure}[t]{0.32\columnwidth}
        \begin{tikzpicture}
        \matrix(ax1){
            \begin{axis}
            [
            ybar,
            xshift=-1cm,
            width=0.42\linewidth,
            height=0.4\linewidth,
      ylabel = { \# Neurons},
      ylabel near ticks,
      xlabel near ticks,
      every minor tick,
      xtick = {0, 0.25, 0.5},
      xticklabels = {0, 0.25, 0.5},
      ytick = {0, 5, ..., 23},
      yticklabels = {0, 5, ..., 23},
    label style={font=\footnotesize},
    tick label style={font=\footnotesize},
      xmin = 0, xmax = 0.52,
      ymin = 0, ymax = 24,
      tick pos=left,
      ymajorgrids,
      grid style={gray!25},
      ytick style={gray!25},
      %axis y line* = right
      legend style={
        at={(0.99, 0.99)},
        anchor=north east,
        draw=none,
        %/tikz/every even column/.append style={column sep=6pt},
        % /tikz/column 2/.style={column sep=10pt},
        % /tikz/column 4/.style={column sep=10pt},
        font={\footnotesize}
      },
            ]
\addplot+ [
        fill,
        draw=none,
        hist={bins=40}] % , data max=0.7,data min=0
        table [x = a, y = a, col sep=comma] {figures/plot2_data_.csv};
\node[anchor=north east] at (rel axis cs:1,1) {\footnotesize Layer 1};
\draw [red] (axis cs:66/256,0)--(axis cs:66/256,24) node [midway,right,yshift=4mm] {\footnotesize 0.26};
% {\small 66/256};
\end{axis}\\
};
%         \end{tikzpicture}
%         % \caption{Layer 1}
%         \label{fig:tau_distr}
%     \end{subfigure}%
% \hfill
% \begin{subfigure}[t]{0.32\columnwidth}
%         \begin{tikzpicture}
\matrix[right=-4mm of ax1](ax2){
            \begin{axis}
            [
            %name=ax2,
            % % position the second `axis` using `at` ...
            % at={(ax1.right of west)},
            % % ... and add a proper `anchor`
            % anchor=west,
            % % optionally/additionally a `yshift` can be stated
            % xshift=2mm,
            ybar,
            width=0.42\linewidth,
            height=0.4\linewidth,
      ylabel near ticks,
      xlabel near ticks,
      xtick = {0,0.3,0.6},
      xticklabels = {0,0.3,0.6},
      label style={font=\footnotesize},
    tick label style={font=\footnotesize},
      ytick = {0, 5, ..., 23},
      yticklabels = {,,},
      xmin = 0, xmax = 0.7,
      ymin = 0, ymax = 24,
      tick pos=left,
      ymajorgrids,
      grid style={gray!25},
      ytick style={gray!25},
      %axis y line* = right
      legend style={
        at={(0.99, 0.99)},
        anchor=north east,
        draw=none,
        %/tikz/every even column/.append style={column sep=6pt},
        % /tikz/column 2/.style={column sep=10pt},
        % /tikz/column 4/.style={column sep=10pt},
        font={\footnotesize}
      },
            ]
\addplot+ [
        fill,
        draw=none,
        hist={bins=40}]
        table [x = b, y = b, col sep=comma] {figures/plot2_data_.csv};
\node[anchor=north east] at (rel axis cs:1,1) {\footnotesize Layer 2};
\draw [red] (axis cs:62/256,0)--(axis cs:62/256,24) node [midway,right,yshift=4mm] {\footnotesize 0.24};
% {\small 62/256};
\end{axis}\\
};
%         \end{tikzpicture}
%         % \caption{Layer 2}
%     \end{subfigure}%
% \hfill
% \begin{subfigure}[t]{0.32\columnwidth}
%         \begin{tikzpicture}
\matrix[right=-2mm of ax2](ax3){
            \begin{axis}
            [
            ybar,
            % % position the second `axis` using `at` ...
            % at={(ax2.right west)},
            % % ... and add a proper `anchor`
            % anchor=west,
            % % optionally/additionally a `yshift` can be stated
            % xshift=2mm,
            %bar width=2pt,
            width=0.42\linewidth,
            height=0.4\linewidth,
      ylabel near ticks,
      xlabel near ticks,
      xtick = {0,0.1,0.2},
      xticklabels = {0,0.1,0.2},
      ytick = {0, 5, ..., 23},
      label style={font=\footnotesize},
    tick label style={font=\footnotesize},
      yticklabels = {,,},
      xmin = 0, xmax = 0.25,
      ymin = 0, ymax = 24,
      tick pos=left,
      ymajorgrids,
      grid style={gray!25},
      ytick style={gray!25},
      %axis y line* = right
      legend style={
        at={(0.99, 0.99)},
        anchor=north east,
        draw=none,
        %/tikz/every even column/.append style={column sep=6pt},
        % /tikz/column 2/.style={column sep=10pt},
        % /tikz/column 4/.style={column sep=10pt},
        font={\footnotesize}
      },
            ]
\addplot+ [
        fill,
        draw=none,
        hist={bins=40}]
        table [x = c, y = c, col sep=comma] {figures/plot2_data_.csv};
\node[anchor=north east] at (rel axis cs:1,1) {\footnotesize Layer 3};
\draw [red] (axis cs:22/256,0)--(axis cs:22/256,24) node [midway,right,yshift=4mm] {\footnotesize 0.09};
% {\small 22/256};
\end{axis}\\
};
        \end{tikzpicture}
        % \caption{Layer 3}
    % \end{subfigure}%
\vspace{-10pt}
\centering
\caption{Spike rate for three spiking layers. Red line shows the average spike rate}
\label{fig:spike_rate}
\vspace{-18pt}
\end{figure}

\vspace{-10pt}
\subsection{Computation complexity analysis}
\begin{figure}[t]
\centering
%\vspace{-9.5pt}

% \begin{subfigure}[t]{0.49\linewidth}
% \resizebox{\linewidth}{!}{
\begin{tikzpicture}
% \pgfplotsset{compat=newest}
% \tikzset{}% rotate the legend images, 
    \begin{axis}[
    name=axis1,
    height=0.4\linewidth,
    width=0.5\linewidth,
      grid,
      grid style={gray!25},
      ytick style={gray!25},
    %scale only axis, 
    xmin=0,xmax=52, 
    xtick={0,10, ..., 52},
    xticklabels={0,10,...,52},
    ymin=0,ymax=1,
    tick pos=left,
    xlabel={\footnotesize Simulation steps},
    ylabel={\footnotesize Val. Correlation},
    label style={font=\footnotesize},
    tick label style={font=\footnotesize},
    legend style={
    draw=none,
    font={\footnotesize}
      },
      ]
    \addplot[line width=1pt,solid,color=blue] 
    table[x=t,y=a,col sep=comma]{figures/plot6_data.csv};
    
    \draw[thin, red!80] (axis cs:0, 0.741761137424304) -- (axis cs:19, 0.741761137424304);
    \draw[thin, red!80] (axis cs:19, 0.741761137424304) -- (axis cs:19, 0);
    
    \node[anchor=north west] (sCL) at (axis cs:21,0.6) {{\renewcommand{\arraystretch}{0.6}\begin{tabular}{c}\footnotesize max at \\ \footnotesize (19, 0.74) \end{tabular}}};
    \node (dCL) at (axis cs: 19,0.741761137424304){};
    \draw[->](sCL)--(dCL);
    
\end{axis}
\end{tikzpicture}
%   }
% \end{subfigure}
% %\hfill
% \begin{subfigure}[]{0.49\linewidth}
%
% %\includegraphics[width = \textwidth]{figures/Total_spike_count.png}
%
\begin{tikzpicture}

\begin{axis}[
    height=0.4\linewidth,
    width=0.5\linewidth,
      grid,
      grid style={gray!25},
      ytick style={gray!25},
    xlabel = {\footnotesize Simulation steps},
    ylabel = {\footnotesize Spikes},
    label style={font=\footnotesize},
    tick label style={font=\footnotesize},
    xmin=0,
    xtick={0,10, ..., 52},
    xticklabels={0,10,...,52},
    ymin=0,
    ytick={0,2500,5000,7500,10000},
    tick pos=left,
    legend style={
    draw=none,
    font={\footnotesize}
      },
    ]
\addplot[line width=1pt] table[x=t,y=avg,col sep=comma] {figures/plot_spike_total_.csv};

\addplot [name path=upper,draw=none] table[x=t,y expr=\thisrow{avg}+\thisrow{dev}, col sep=comma] {figures/plot_spike_total_.csv};
\addplot [name path=lower,draw=none] table[x=t,y expr=\thisrow{avg}-\thisrow{dev}, col sep=comma] {figures/plot_spike_total_.csv};
\addplot [fill=blue!10] fill between[of=upper and lower];

\draw[thin, red!80] (axis cs:0, 3451.03615404768) -- (axis cs:19, 3451.03615404768);
\draw[thin, red!80] (axis cs:19, 3451.03615404768) -- (axis cs:19, 0);

\node[anchor=south] (sCL) at (axis cs:16,6000) {{\renewcommand{\arraystretch}{0.6}\begin{tabular}{c}\footnotesize 3451 spikes\\ \footnotesize at 19 steps \end{tabular}}};
    \node (dCL) at (axis cs: 19,3451.03615404768){};
    \draw[->](sCL)--(dCL);

\end{axis}

\end{tikzpicture}
% }
% \end{subfigure}
\vspace{-5pt}
\caption{ANN converted SNN.}
\label{fig:ann_2_snn}
\vspace{-18pt}
\end{figure}
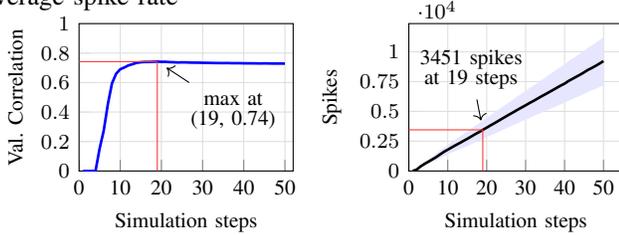

% \vspace{-3pt}
SNNs are appealing for their potentials in achieving efficient hardware implementation \cite{10.3389/fnins.2018.00774}. One primary source of efficiency is sparsity. We evaluated the spike rate for neurons in different layers and average spike count in each layer to quantify the sparsity in our network. The results are presented in Fig.~\ref{fig:spike_rate}.
Most of the neurons in the all layers have spiking probability well below $50\%$. 
%The probability for most of the neurons to spike in all the layers is well below $50\%$. 
The average spike rates are $26\%$, $24\%$, $9\%$ for the three spiking layers respectively. In each inference, out of the 770 neurons in the whole network, there are 150 spikes generated on average.
% On average, in each inference, there are 66 spikes in the first layer, 62 in the second, and 22 in the third while there are 256 neurons in each layer. In total, there are 150 spikes in each inference.
Due to the rate approximation, the SNN converted from ANN requires multiple simulation steps on the same input for one output to achieve sufficient accuracy, in this case 19 steps for one prediction. In total, around 3000 spikes are used per inference as shown in Fig.~\ref{fig:ann_2_snn}. The ANN-converted SNN uses 20 times more spikes than the proposed one, indicating longer latency and higher energy consumption in hardware implementation.

% for the plot on the right with shaded area: 
% https://tex.stackexchange.com/questions/430650/error-bars-as-shaded-area

In conventional ANNs, the weighted sum computation for each input to a neuron requires a MAC operation. Whereas in the proposed SNN, as the spike status is either 1 or 0, this process has been replaced by addition operations, which requires much less power~\cite{horowitz_11_2014}. Here, we conservatively assume three additions as one MAC operation according to the ratio of the energy cost for floating point operations reported in~\cite{horowitz_11_2014} for the comparison of the total number of operations. To have a fair comparison between the proposed SNN and ANN, it is important to remark that the SNN needs to update its membrane potentials as described in \eqref{eq:mem_upd} once per inference, this amounts to an additional MAC operation per each neuron. In this comparison, three memory loads and one store are assumed for each MAC while two loads and one store are assumed for each addition.
% In this comparison, two memory load operations and one memory store operation are assumed for each computation. 
The average spike rates in Fig.~\ref{fig:spike_rate} are used for the analysis. Additions and the corresponding memory accesses are not executed when there is no spike. The results are summarized in Table~\ref{tab:comp}.
Thanks to enhanced STBP training and the high level of sparsity, the proposed SNN requires $6.8\%$ operations and $9.4\%$ memory access compared to the ANN while achieving same level of accuracy.
% Specifically, in our case, the SNN performs 487K less operations and 1414K less memory access than the ANN.
% and an equal amount of operation between the SNN and the ANN implementation would be reached for an average spike rate of XYZ of the SNN.

% \begin{table}[htbp]
% \caption{Computation Comparison}
% \begin{center}
% \begin{tabular}{|c|c|c|}
% \hline
% \textbf{Operation}&\multicolumn{2}{|c|}{\textbf{Decoder}} \\
% \cline{2-3} 
% \textbf{} & \textbf{\textit{ANN} \cite{willsey_real-time_2021}} & SNN\\
% \hline
% MAC & 529K & 25K  \\
% \hline
% ADD & - & 59K  \\
% \hline
% Mem & 1,588K & 253K \\
% \hline
% % \hline
% % MAC & 529408 $^{\mathrm{a}}$& 25346  \\
% % \hline
% % ADD & - $^{\mathrm{a}}$& 59073  \\
% % \hline
% % Mem & 1,588,224 & 253,257 \\
% % \hline

% \end{tabular}
% \label{tab:comp}
% \end{center}
% \end{table}

\vspace{-3pt}
\section{Conclusion}
\vspace{-3pt}
In this paper, we present a spiking neural network and its training method to decode continuous-valued finger velocities for implantable BMI applications. The proposed network is trained with STBP backpropagation enhanced by trainable decay factor and reset-by-subtract techniques to improve the accuracy while keeping low computation complexity.
% The proposed SNN improves its performance by incorporating trainable decay factor and reset-by-subtract techniques. The network is trained with backpropagation-based approach. 
We compared the performance of the proposed SNN, Kalman filter, ANN, and SNN converted from ANN on two datasets for open-loop finger decoding tasks. The proposed SNN achieves the same level of correlation coefficient as the state-of-the-art decoders, while showing significantly less spike count than SNN converted from ANN and $6.8\%$ operations and $9.4\%$ memory access compared to the ANN decoder, indicating potential in achieving energy-efficient hardware implementation. %Quantization and close-loop control performance remain to be investigated in the future.

\bibliographystyle{IEEEtran}
\bibliography{main.bib,bstctl}

\end{document}